\def\year{2015}
\documentclass[letterpaper]{article}
\usepackage{aaai}
\usepackage{times}
\usepackage{graphicx}
\usepackage{float}
\usepackage{subfigure}
\begin{document}

\title{Maestoso: An Intelligent Educational \\Sketching Tool for Learning Music Theory}
\author{Paul Taele\\
Sketch Recognition Lab\\
Texas A\&M University\\
College Station, TX, USA 77843\\
ptaele@cse.tamu.edu\\
\And
Laura Barreto\\
Computer Science Department\\
Vassar College\\
Poughkeepsie, NY, USA 12604\\
labarreto@vassar.edu\\
\And
Tracy Hammond\\
Sketch Recognition Lab\\
Texas A\&M University\\
College Station, TX, USA 77843\\
hammond@cse.tamu.edu\\
}
\maketitle

\begin{abstract}
\begin{quote}
Learning music theory not only has practical benefits for musicians to write, perform, understand, and express music better, but also for both non-musicians to improve critical thinking, math analytical skills, and music appreciation. However, current external tools applicable for learning music theory through writing when human instruction is unavailable are either limited in feedback, lacking a written modality, or assuming already strong familiarity of music theory concepts. In this paper, we describe Maestoso, an educational tool for novice learners to learn music theory through sketching practice of quizzed music structures. Maestoso first automatically recognizes students' sketched input of quizzed concepts, then relies on existing sketch and gesture recognition techniques to automatically recognize the input, and finally generates instructor-emulated feedback. From our evaluations, we demonstrate that Maestoso performs reasonably well on recognizing music structure elements and that novice students can comfortably grasp introductory music theory in a single session.
\end{quote}
\end{abstract}

\section{Introduction}
People interested in music and aspiring to become musicians will greatly benefit from studying music theory to comprehend its structure. Such aspiring musicians can see the practical benefits of its mastery including more accurately writing and performing music compositions, better understanding the craft through reading and listening to existing works, and more fully expressing their ideas from its components~\cite{Miller:2005:Book:MusicTheory}. However, the advantages of learning music theory are not solely limited to improving musical skill. Non-musician listeners can also appreciate its practical benefits such as better fostering critical thinking~\cite{Telegraph:2012:Web:Music}, more enthusiastically improving math analytical skills through an interesting real-world domain~\cite{Topoglu:2014:Procedia:MusicEducation}, and having greater appreciation of music that they enjoy~\cite{Nelson:2014:Book:Music}.

A common scenario in traditional music theory instruction involves students receiving in-class guidance on newly-introduced concepts from a music expert instructor, practicing their writing of these concepts, and receiving critique and feedback from the instructor on their written input. On the other hand, while this form of instruction may be optimal for students to learn effectively, in reality access to valuable personalized instruction from human instructors in different subjects on their learning progress is not always readily available when the student desires, especially for music theory. As a result, students may turn to self-study resource options to supplement their in-class learning. However, conventional textbooks and workbooks heavily relies on rote memorization while lacking dynamic feedback. Current computer-assisted music educational interfaces are also constrained on the types of feedback they produce and dominantly lack a sketching modality and automatic recognition of sketched input. Furthermore, prevalent music notation software assumes that users have strong existing music theory knowledge and are geared more heavily towards advanced users to produce professional music compositions instead of novice users learning music fundamentals.

\begin{figure}[ht]
	\centering
	\subfigure[]{
		\includegraphics[width=.9\linewidth]{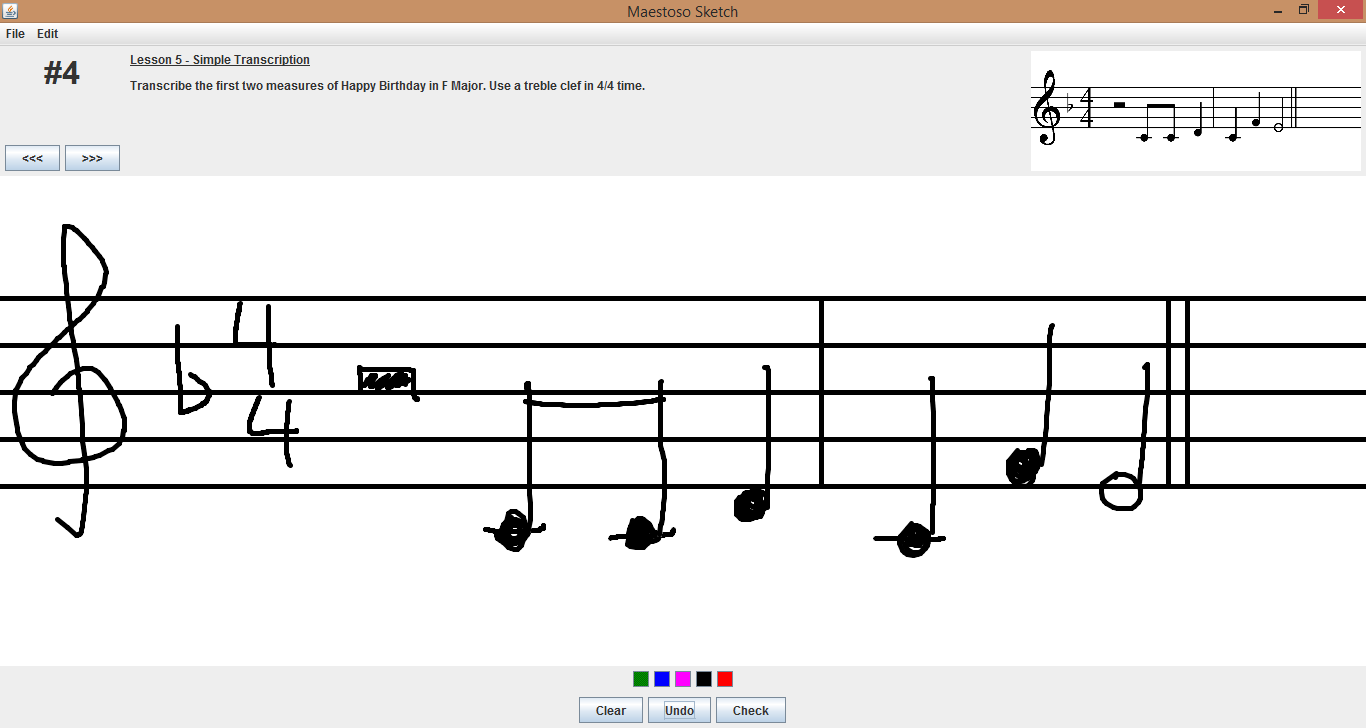}
		\label{fig:interface:raw}
	 }

	 \subfigure[]{
		\includegraphics[width=.9\linewidth]{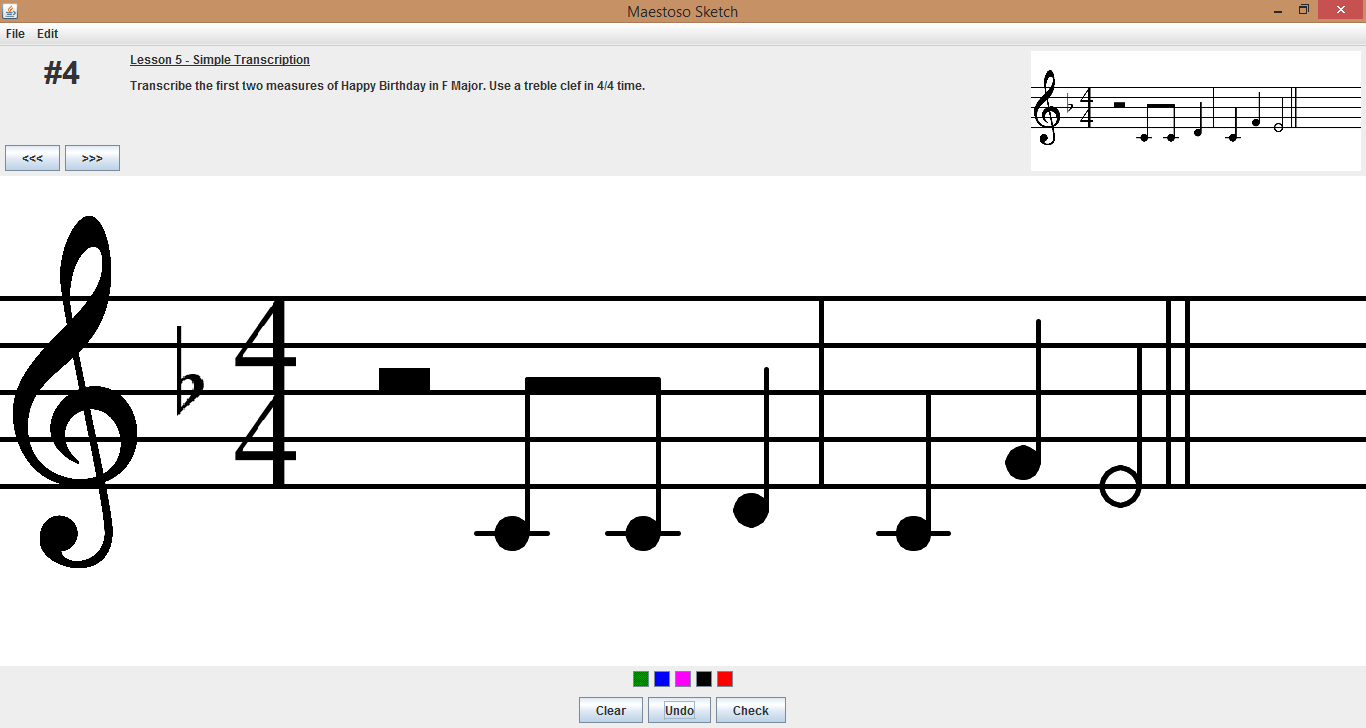}
		\label{fig:interface:clean}
	 }

	\caption{A sample screenshot of the Maetoso interface: (a) with the original raw strokes displayed, and (b) with the generated beautified strokes displayed.}
	\label{fig:interface}
\end{figure}

In this paper, we introduce Maestoso, an intelligent computer-assisted educational application that allows people to learn music theory through sketching, and enabling them to progress through provided lessons of important fundamentals in music theory (Figure~\ref{fig:interface}). Users not only can directly quiz their sketching knowledge of music structure, but can also practice their sketched responses to prompted questions with accompanying text hints and visualized solutions. The interface also responds to users with instructor-emulated feedback and critique of their written music structure input by identifying both corrections to their solution and also specialized feedback where students were unsuccessful in their solutions. From our evaluation of the sketch recognition and educational interface capabilities of Maestoso, we discovered that the interface was robust to novice users' sketching styles while also allowing them to successfully learn music theory concepts during their usage sessions.

\section{Related Work}
Since Maesotoso is an intelligent sketch-based music educational tool, related works include computer-assisted educational interfaces, music notation software, and music recognition techniques.

\subsection{Music Education Applications}
The growing affordability of mobile touch devices and the rising acceptance of digital educational resources are further encouraging developers and educators to expand beyond traditional computer-assisted educational tools for a variety of subjects such as music. For the former, music education apps have ranged from those that provide mobility convenience in reviewing music theory concepts for older users such as Tenuto~\cite{Tenuto:2014:App:Tenuto} and Theory Lessons~\cite{TheoryLessons:2014:App:TheoryLessons}, to those that focus on presenting novel ways to express music for younger users such as Paint Melody \cite{PaintMelody:2014:App:PaintMelody}. For the latter, massively open online courses (MOOCs) such as Coursera~\cite{Coursera:2014:Web:Music} and textbook digital media such as Foundations of Music~\cite{Nelson:2014:Book:Music} provide users with educational resources to pursue formal music theory study outside the traditional classroom. However, these applications and services primarily provide at most general binary feedback for testing users on their studied music theory concepts, and also lack a sketching modality for users to receive automated feedback on their written input.

\subsection{Music Notation Applications}
On the other side of the spectrum are more dedicated tools for composing and notating music. Established music notation tools such as Finale~\cite{Finale:2014:App:Finale} and Sibelius~\cite{Sibelius:2014:App:Sibelius} provide rich features for composing music using a mouse, microphone, scanner, or musical keyboard, and Finale also incorporates optional educational functionalities such as flashcards. Other types of similar applications make use of pen and touch modalities for notating music, such as early system Music Notepad~\cite{Forsberg:1998:UIST:MusicNotepad} for penning music notation, recent app NotateMe~\cite{NotateMe:2013:App:NotateMe} for touch gesturing music notation, and also MusicReader~\cite{Leone:2008:MindTrek:MusicReader} for penning basic music annotation. Works such as MusicFlow~\cite{Tan:2010:SPIE:MusicFlow} and SoundBrush~\cite{Brooke:2013:Web:SoundBrush} also incorporate multi-touch for traditional and unique takes on notating music. However, these tools largely assume that users already have strong familiarity of music theory for writing music, and lack features for effectively studying music theory.

\subsection{Intelligent Sketch-enabled Education Applications}
As sketching continues to play an important role in the learning process within traditional classrooms, intelligent computer-assisted educational systems have similarly incorporated sketching modalities for a variety of academic subjects. These systems include but are limited to Physicsbook~\cite{Cheema:2012:IUI:PhysicsBook} for physics, Hashigo~\cite{Taele:2009:IAAI:Hashigo} for written Japanese, iCanDraw~\cite{Dixon:2010:CHI:iCanDraw} for figure drawing, MathPad2~\cite{LaViola:2004:SIGGRAPH:MathPad2} for mathematics, and BiologySketch~\cite{Taele:2009:IUISRW:Biology} for biology. In developing Maestoso, we have taken the successful lessons from these previous systems that were most appropriate for the domain of music theory.

\subsection{Written Music Recognition}
Research work for automatically recognizing music expression is well-explored and active in numerous directions, including at the acoustic (e.g.,~\cite{Ramirezand:2006:AAAI:Evolutionary}) and emotional (e.g.,~\cite{Wu:2014:MM:MusicEmotion}) level. For printed and handwritten music, most efforts have focused on existing optical music recognition (OMR) techniques~\cite{Rebelo:2012:IJMIR:OMR}. Representative efforts include k-nearest neighbor-based OMRs for handwritten scores~\cite{Rebelo:2011:ICMLA:MetricLearning}, writer identification and staff removal~\cite{Fornes:2012:IJDAR:GroundTruth}, hidden Markov-based OMRs for notation input~\cite{Lee:2010:CAMP:HMM}, and conventional OMRs and image processing techniques for automatically recognizing and playing back simple handwritten music notation~\cite{Baba:2012:SIGGRAPH:Gocen,Yamamoto:2011:UIST:onNote}. However, these approaches generally either focus on recognizing handwritten music on paper or on recognizing stylus input from advanced users, which differ from the recognition approaches that Maestoso uses for its target audience of novice users with a stylus.

\section{Interaction Design Process}
\begin{figure}
	\centering
	\includegraphics[width=.8\linewidth]{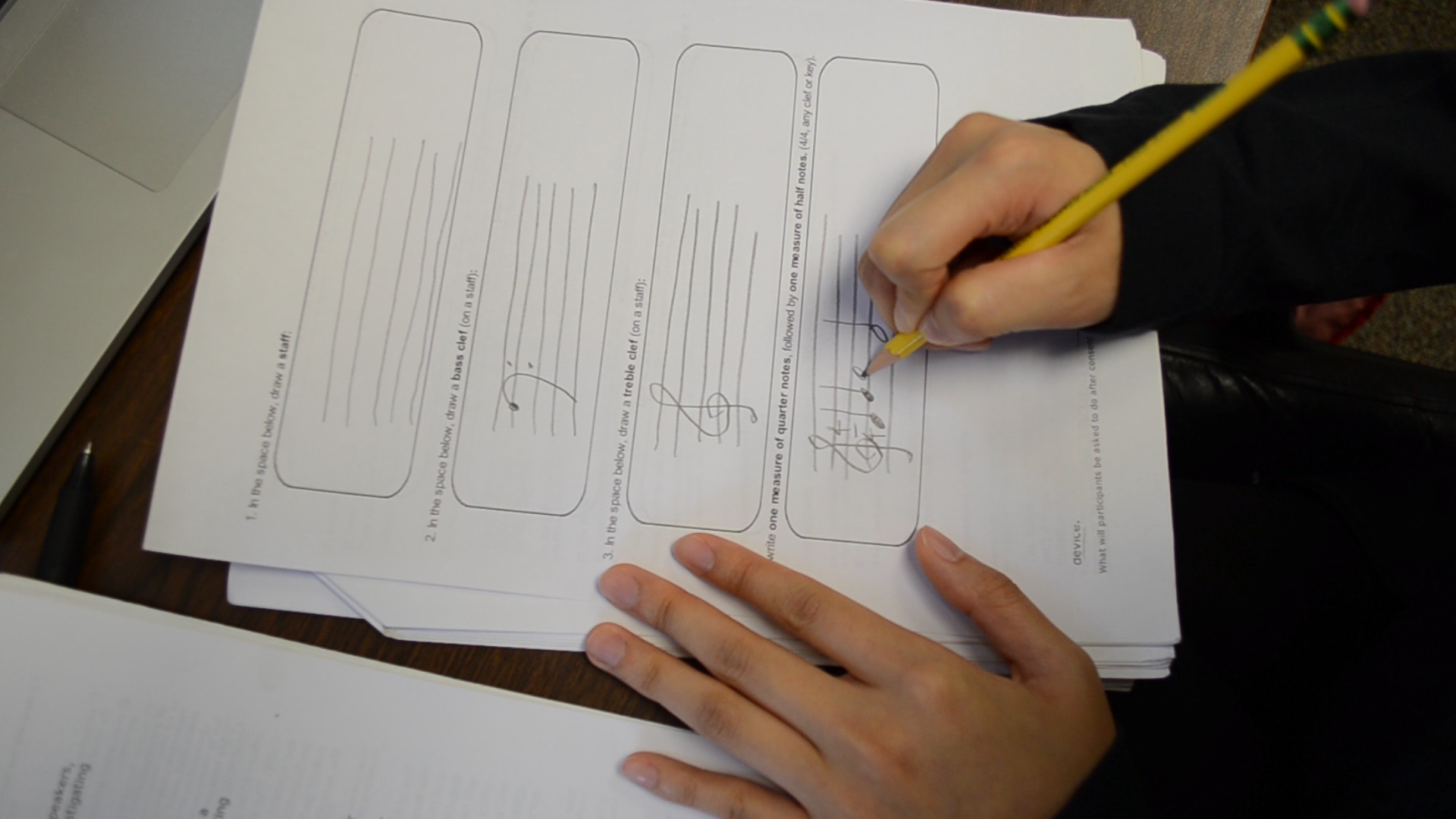}
	\caption{A user with novice music theory knowledge writing music component responses.}
	\label{fig:paper:writing}
\end{figure}

Prior to the development of Maestoso, we first sought out individuals with only passing music knowledge to better understand general drawing behaviors of novice students for music recognition.  Through convenience sampling, we recruited six individuals -- between ages 21 and 32, two females -- whom were all non-musicians and also self-reported having limited music theory knowledge. From these six users, we provided them with foundational music theory questions on paper from texts such as~\cite{Nelson:2014:Book:Music} for drawing individual musical components (e.g., notes, clefs, staff). We then asked them to pencil their responses while video recording only their responses put on paper (Figure~\ref{fig:paper:writing}), similar to the user studies for observing drawings from~\cite{vanSommers:2009:Book:Drawing}. Whenever a participant did not know how to draw a particular musical component, we would briefly show them by demonstration once so that they were aware of the response but not heavily influenced by the study conductor's drawing style. The main sketching patterns from all participants with limited music theory knowledge included musical components drawn more carefully and in separate strokes (e.g., a note's parts draw in several strokes instead of merged as one). As a result, we used these insights for designing our music recognition on novice students' writing behaviors.

\section{Music Recognition System}
\begin{figure}
	\centering
	\includegraphics[width=1\linewidth]{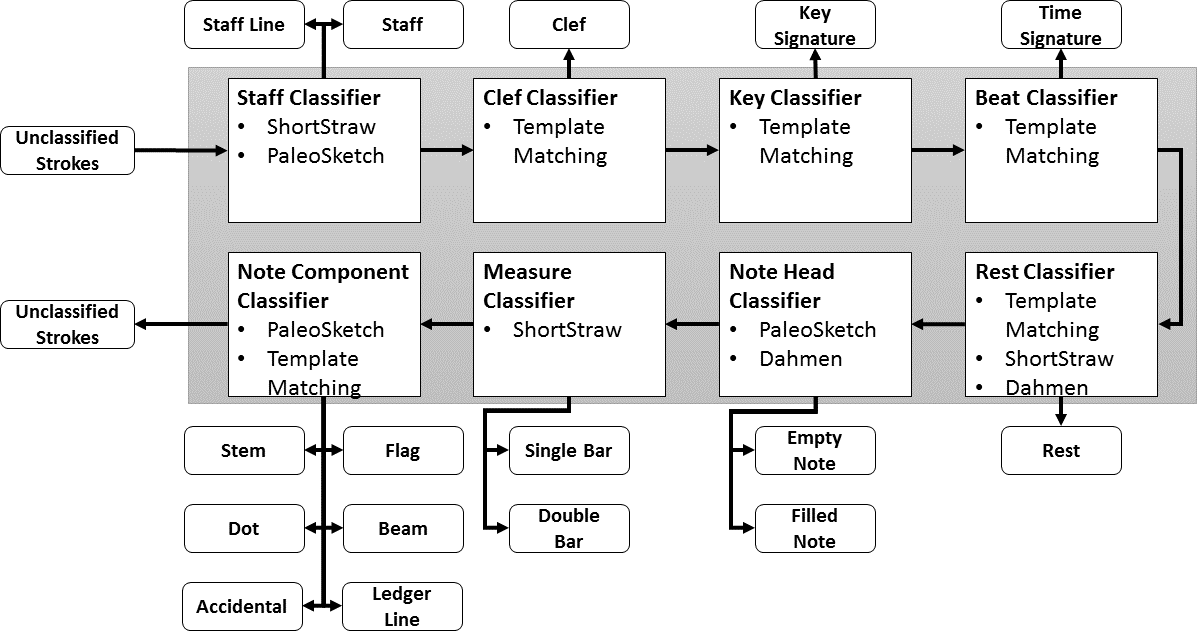}
	\caption{Hierarchy of classifiers used in Maestoso, in order of execution.}
	\label{fig:classifier:hierarchy}
\end{figure}

One of the main goals for developing the music recognition aspect of Maestoso was to ensure that recognition was robust for novice users' written music input, while also ensuring that their input was not being recognized for incorrect or sloppy writing practices. That was one of the motivations for avoiding existing OMR techniques, since those approaches were designed more for greater accuracy for advanced users' written input instead of more proper form for novice users' written input. As a result, we instead utilized a collection of sketch recognition techniques to build a hierarchy of classifiers, in order to support music theory concepts that are important for novice users to initially learn. We provide a hierarchy of our system in Figure~\ref{fig:classifier:hierarchy} and elaborate further on each of these hierarchical components in the following sections.

\subsection{Template Matching}
Some common music symbols such as staves, notes, and longer duration rests (e.g., half and whole rests) possess easily-defined geometric structures, and would perform well with existing geometric-based sketch recognition techniques described later in the paper. However, for other common music symbols with more visually complex structures such as clefs, accidentals, and shorter duration rests (e.g., quarter and eighth rests), we instead turn to insights from both touch gesture recognizers and sketch corner finders. The general template matching approach that we employed for these types of music symbols first involved uniformly transforming the raw sketch strokes' points using ShortStraw's stroke transformation steps~\cite{Wolin:2008:SBIM:ShortStraw}, specifically resampling to 64 points, scaling to 250 pixels, and translating to (0,0) in Cartesian coordinates. We identically performed this transformation when we preprocessed the music symbol classes, where each class contained twenty separate user-drawn candidate templates for matching.

Afterwards, we similarly followed the template matching steps of touch gesture recognizers such as \$1~\cite{Wobbrock:2007:UIST:Dollar} and Protractor~\cite{Li:2010:CHI:Protractor}, where we match the sketched input to each music symbols' candidate templates. Since music symbols can be drawn with more than one stroke (e.g., bass clef, accidental sharp), we also combine strokes of sketched input similarly to the multi-stroke template matching approach used in \$N-Protractor~\cite{Anthony:2012:GI:NDollar}.

One modification in our template matching approach for sketched music symbols from the prior gesture recognizers is that we compare the sketched input to a candidate template by calculating the Hausdorff distance between their set of points, instead of summing the Euclidean distances of their corresponding one-to-one points. We observed that this modification improved classification of sketched music symbols, and also performed similarly well in other sketching domains such as sketched engineering diagrams for instructional interfaces~\cite{Valentine:2011:IUISRW:ShapeComparison}. We subsequently refined the similarity score calculation -- previously used in touch gesture recognizers such as \$1 -- specifically for Hausdorff distance, in order to provide more well-defined similarity confidence values between the sketched strokes and the candidate template below:

\begin{center}
$score = 1 - \frac{|\frac{1-d}{\sqrt{S^2+S^2}}|}{10}$
\end{center}

Once we locate the template with the highest score, our template matching approach returns both the predicted shape and its corresponding similarity score that some of the music symbol classifiers utilize in Maestoso.

\subsection{Staff Classification}
The staff, which consists of five parallel horizontal staff lines, plays an important role in setting the placement of the rest of the music symbols. For our staff classifier, we first take cues from the line tests in PaleoSketch~\cite{Paulson:2008:IUI:PaleoSketch} and ShortStraw~\cite{Wolin:2008:SBIM:ShortStraw} for handling each of the five individual staff lines. Specifically, since staves in Maestoso span the entire width of the drawing space, we calculate the entire path distance of the sketched stroke between its endpoints, and then calculate the ratio to the drawing space's width. We then measure the incline angle that is formed from the Euclidean line between those two endpoints. Finally, if the sketched stroke is greater than 95\% for the line test and less than 5\% for the angle test, it is then classified as a staff line.

Once the user has drawn five classified staff lines, the staff classifier first proceeds to classify those lines as a staff, and then beautifies the staff lines by straightening the top and bottom staff lines while evenly spacing the three remaining staff lines in-between. From this information, the staff classifier finally calculates the vertical distance of the space between two consecutive staff lines (i.e., the step value) and enumerates the staff lines and spaces of the staff. The step value from the former helps with classifying the relative size of the music symbols, while the enumerations from the latter help determine the correctness of the music symbols' placement relative to the staff.

\subsection{Clef, Key, and Beat Classification}
If the sketched strokes are not recognized by the staff classifier, they will then proceed to the next group of classifiers for handling clefs, keys, and beats. These corresponding classifiers together perform similarly for classifying the sketched strokes by first template matching to music symbols of that type (i.e., the matching check). The clef classifier does so for the treble and bass clef, which are the most commonly used clefs in written music and essential learning for novice students; the key classifier does so for sharp and flat accidentals that make up the key signature; and the beat classifier does so for digits ranging from 0 to 9 that make up the time signature.

These classifiers then check successful template matches for the correctness of their height, position, and stroke count based on their music theory definitions (i.e., the definition check). As an example for the clef classifier, sketched strokes are classified as a bass clef if it has a template matching similarity score above 0.85, is no greater than four steps in height, consists of three strokes, and spans approximately the top three-thirds of the staff. The definition check is straightforward for the key classifier, while the beat classifier also checks for two digits that are vertically aligned and spanning the staff's height to complete a time signature.

\subsection{Rest, Note Head, and Measure Classification}
Sketched strokes that are still unclassified then proceed to the next group of classifiers consisting of rests, note heads, and measure bars. The rest classifier initially checks the sketched strokes with template matching for music symbols that are not easily geometric-definable (e.g., quarter and eighth rests). Otherwise, the rest classifier first checks if the sketched stroke contains a partial rectangular outline for geometric-definable rests (e.g., half and whole rests) using line and corner information from ShortStraw, and then checks for a filled rectangle using a simplified version of the closed shape stroke density test from ~\cite{Dahmen:2008:AAAI:Scribbles} that checks if at least 80\% of a stroke is contained inside the rest's rectangular outline. If either matching check succeeds, then we perform a definition check on the sketched strokes.

With the note head classifier, sketched strokes are instead checked for whether they are a filled or empty note head by taking cues from PaleoSketch's circle test. Specifically, we first check if the sketched strokes' path distance is approximately equal to the circumference of the ideal circle formed around its radius, then check if their bounding box's width and height have similar lengths, and finally check if the sketched strokes form a closed shape. If so, we again use the closed shape stroke density test to determine if the note head is empty or filled.

The simplest of the three classifiers is the measure classifier, which checks for single and double measure bars using ShortStraw's line test, and by determining if the stroke is angled within 5\% of a vertical line and spans the vertical length of the staff. For double measure bars, we additionally check that the distance between adjacent bars is within half a step.

\subsection{Note Components Classification}
The last stage of classification focuses on the remaining note components that are the stems, flags, dots, beams, accidentals, and ledger lines. For each note component, we first determine if the component is within half a step of a specific area of a note head (e.g., stems, dots, ledger lines, accidentals) or stem (e.g., beams, flags). We then rely on various geometric shape tests from PaleoSketch to classify some of the note components, such as a line test for stems, beams, and ledger lines. For dots, we employ a naive dot test that checks if a stroke contains at most two points, while flags are approximated as ascending or descending lines and whose lengths are between a diagonal line and two lines at right angles, in order to handle drawing variations in flags. For accidentals, we rely solely on template matching. If the sketched strokes remain unclassified after performing a definition check, then we leave the strokes unclassified for additional context from subsequent strokes.

\section{Learning Interaction System}
\begin{figure}
	\centering
	\includegraphics[width=.8\linewidth]{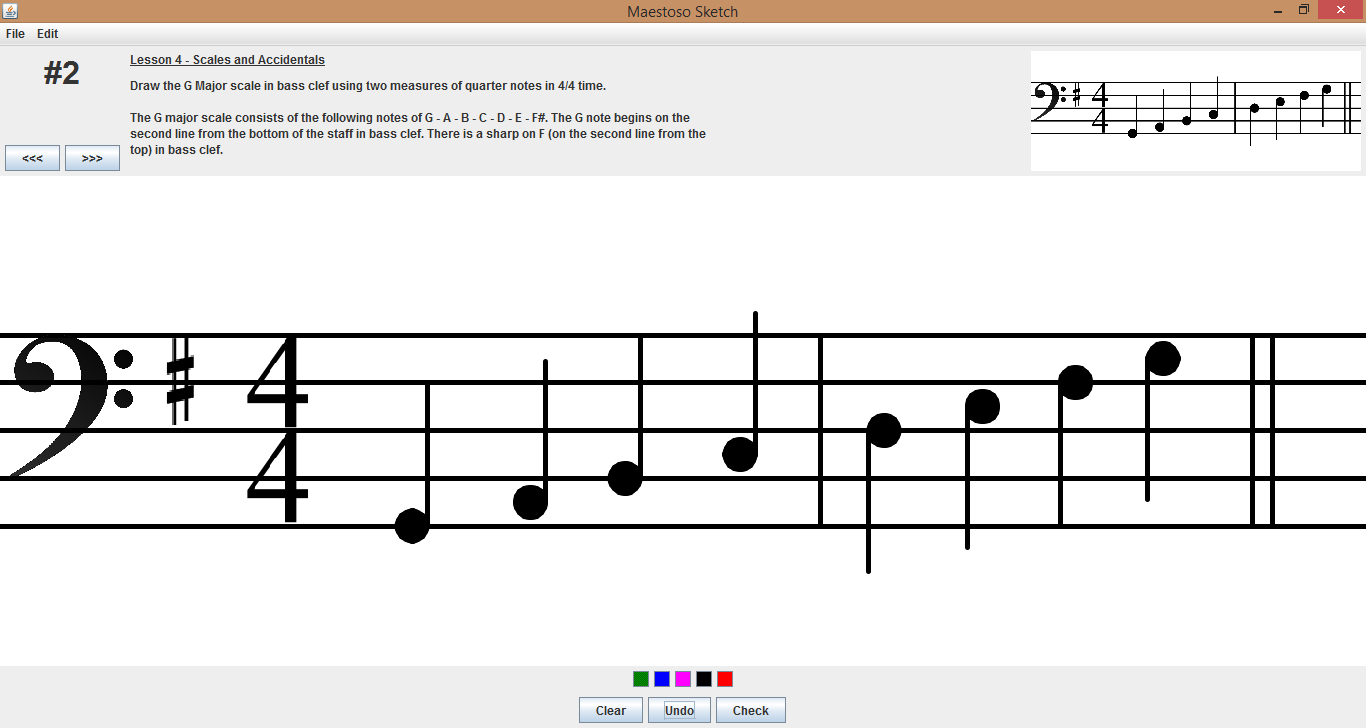}
	\caption{Interface for practicing on prompted music theory questions. The quiz interface removes the text hints and graphic solution.}
	\label{fig:practice:mode}
\end{figure}

Built on top of the music recognition system is the learning interaction interface, where users sketch their music structure responses and receive automated feedback to better understand music concepts. The main sketching interface is a window with three parts: the top lesson area, the middle sketching area, and the bottom interaction area (Figure~\ref{fig:practice:mode}). The top lesson area contains the question number, question text, accompanying text hint, solution image, and buttons to navigate between the different questions in practice mode, while the hint, image, and back navigation button are disabled in quiz mode. The middle sketching area is straightforward and consists of a space for users to sketch their answers. By default, the user's input is automatically beautified to recognized music components, and can be disabled in the menu bar. The bottom interaction area is similarly straightforward, containing a color pallet for the current stylus ink, as well as buttons to clear, undo, and check answer.

\begin{figure}
	\centering
	\includegraphics[width=.8\linewidth]{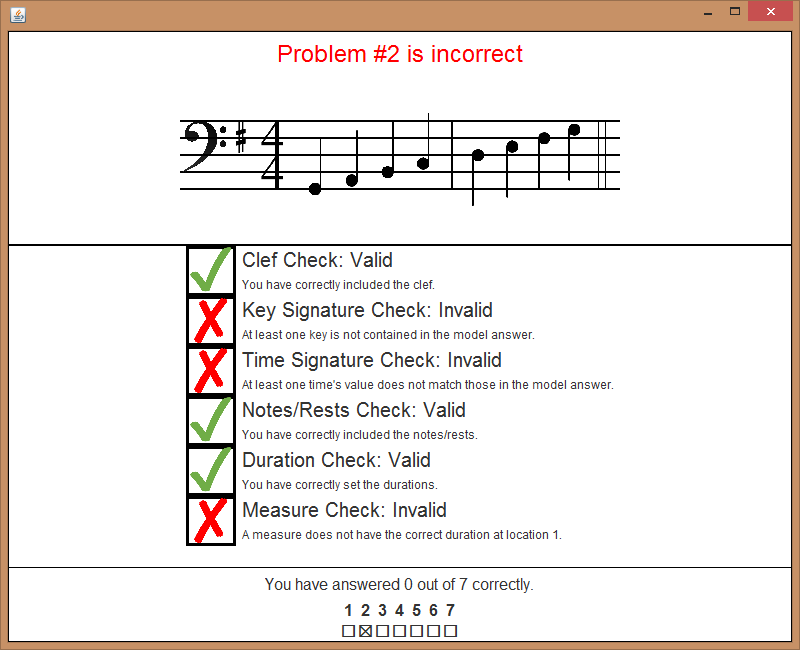}
	\caption{Output window displaying personalized feedback to users of the indivdiual question feedback.}
	\label{fig:feedback:window}
\end{figure}

\begin{figure}
	\centering
	\includegraphics[width=.8\linewidth]{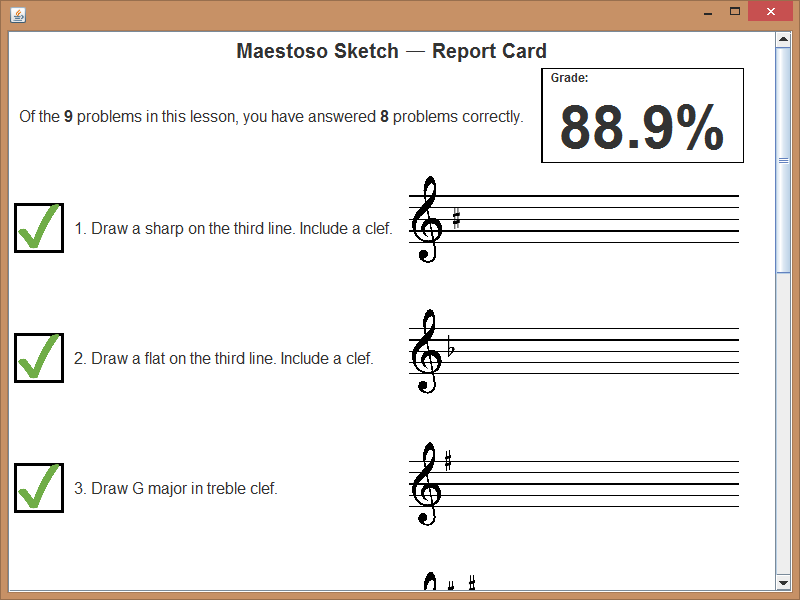}
	\caption{Output window displaying personalized feedback to users of the entire quiz overall report.}
	\label{fig:report:window}
\end{figure}

In addition to the main sketching interface window, there are two others windows available to the user: the feedback window (Figure~\ref{fig:feedback:window}) and the report window (Figure~\ref{fig:report:window}). The former feedback window appears after the user clicks the check button to check their response to the question, and consists of the following components: a sentence indicating whether the response is correct or not, the model solution image, a set of criteria checks (i.e., staff, clef, key signature, time signature, duration, measure) that are individually enabled depending on the given problem, and a progress of questions answered correctly, incorrectly, and in progress. Each criteria check not only lists the general criteria being checked, but also includes more specific detail related to that criteria. For example, if a user drew a note in the incorrect position or duration, or if the user wrote the incorrect clef or key signature, or if the user included the incorrect number of beats in a particular measure, then the criteria check's detailed feedback will specifically state as such. The latter report window appears after the last question is checked during quiz mode only, which gives an overall score of the quiz, an update of the number of questions answered correctly, as well as a list of the questions that includes the question number and text, the model solution image, and whether the user answered them correctly.

\begin{figure}
	\centering
	\includegraphics[width=.9\linewidth]{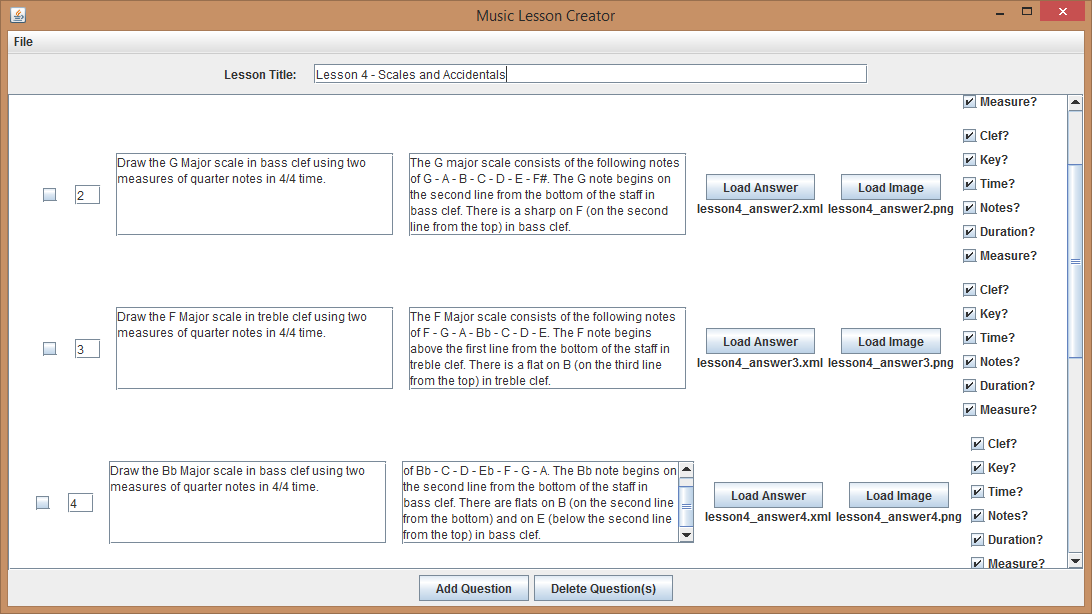}
	\caption{Interface for instructors to create personalized music lessons.}
	\label{fig:lesson:builder}
\end{figure}

In addition to the core Maestoso application, we also included an accompanying lesson builder for instructors to develop lessons specific to their curriculum (Figure~\ref{fig:lesson:builder}). In order to assist in the development of our lesson builder, we consulted with four different music experts for their feedback on appropriate features: an engineering student with years of music study, a professional music instructor with a formal graduate degree in music studies, a part-time music instructor with over a decade of music training, and a music department graduate student from a major university.

Based on our discussions with our consulted experts for building a lesson builder appropriate to Maestoso, we included the following features: editable question numbers for easy rearranging of the questions, a text box for inputting questions, an additional text box for inputting hints, buttons for including answer and image files, and checkboxes for selectable criteria checks used in Maestoso. This last feature was strongly suggested from the music experts as it allowed them to provide greater flexibility on how they can design the questions. For example, an instructor can disable checks on clefs if any clef was fine, disable checks on keys if the solution does not require them, or disable notes if only their durations were requested. The selection of these criteria checks thus determines which classifiers are enabled during the practice and quiz modes in Maestoso.

What is also important to note is that as the instructor is designing the lesson, they may not have files readily available containing the model answer and solution image. The first reason is that we allow the instructor to first sketch the model answer after running the lesson once on the Maestoso interface, then save the answer as an XML file, and finally include it in the lesson builder. The second reason is that the instructor can have flexibility of what kind of graphic to include as the solution image (e.g., screeenshot, handwritten, animation).

\section{Results and Discussion}
\subsection{Recognition Evaluation}
For our first evaluation, we wanted to determine how well the music recognition system performed for novice users' written music input. In order to do that, we made use of the original group of novice users that provided sketches of musical components made on paper, and asked them to similarly provide written input on those components with a stylus. The setup uses a Wacom Bamboo tablet connected to a laptop, and the data set consisted of all the components that were used in the music recognition system, such as staffs, clefs, keys, beats, rests, notes of different variations, measure bars, and so on. We lastly asked the participants to sketch twenty instances of each component in consecutive order while we recorded their sketched input.

Following the completion of these tasks, we then analyzed the performance of the music recognition system by using all-or-nothing recognition~\cite{Wolin:2008:SBIM:ShortStraw} for determining a component as being either completely correct or not at all. What we discovered was that all the users were able to draw the musical components in the data set on average of over 95\%, where the easier components such as measure bars and staves were recognized almost perfectly correct, while more complex components such as clefs and more complex note variants performed greater than 90\%.

\subsection{Interaction Evaluation}
Our second evaluation consisted of having participants run through music lessons in practice mode, and subsequently strive to achieve the highest score in quiz mode after they felt comfortable with that lesson's set of questions. For this evaluation, we recruited a separate group of ten participants -- ages 18 through 27, six females -- whom self-reported that they had only passing knowledge of music theory but had at least some familiarity with using stylus- and touch-enabled computing devices. We then asked them to run through five different lessons of varying difficulty. These lessons were designed to be self-contained for introducing core beginner music theory concepts, and we consulted with our four music experts as well as different music theory textbooks and online music theory websites to define the specific questions of these lessons. These five lessons consisted of the following topics: 1) staffs and clefs, 2) key and time signatures, 3) basic notation, 4) scales and accidentals, and 5) simple transcription.

After instructing the participants to go through the five lessons in both practice and quiz modes, we allowed them to perform the study while we observed and timed their performance and became available if they encountered any issues. In general, the participants were able to complete all five lessons in one session that lasted from forty-five minutes for one user to two hours for another. From the study, we observed that all the users were able to complete the quizzes while making either zero mistakes most of the time and erring on a question on a few occasions. We also discovered that users who spent longer in their session scored slightly higher, and that everyone had varying difficulties once they reached the last lesson, since that lesson tested their memorization skills of transcribing measures of classic well-known songs.

\section{Future Work}
The work in Maestoso has focused primarily on teaching music theory concepts to novice learners, and as such we have built both the music recognition and learning interaction systems to this particular group of users. Therefore, one promising next step for Maestoso is to expand it into incorporating more advanced music theory concepts, so that we can further narrow the gap between educational tools accessible to novice students and professional tools used by advanced users. The reason is to not only provide students with additional practice through an intelligent sketching interface, but to also reduce the barriers and decrease the time for previously novice students to exploit the capabilities of professional students to potentially accelerate their music theory learning.

In addition to broadening Maestoso to users across a wider range of background music theory knowledge, another important next step is to work with local educators in developing appropriate lesson plans with Maestoso for teaching music concepts that are highly relevant and interesting to them. Two examples include employing Maestoso as a tool for better understanding music composition to music students and more concretely grasping real-world arithmetic problems to mathematics students. From the work in developing such lesson plans, we would also like to work with educators in deploying Maestoso and observing its use in different classroom setups with these lesson plans, such as in classrooms that encourage students to bring their own devices, in computer labs where students work with in-school technologies, and outside of class with personal devices for additional interactive self-study or practice of music theory concepts.

\small

\section{Conclusion}
In this paper, we introduced and described our Maestoso application, a computer-assisted sketching interface for music theory instruction. The application relies on a collection of sketch recognition approaches for recognizing novice students' written music, and employs a sketching interface and content partly driven by music expert feedback. From our evaluations, we discovered that the music recognition system performs reasonably well on music components, and that novice users were able to similarly perform reasonably well when quizzed on fundamental music theory concepts.

\section{Acknowledgements}
We would like to thank the DREU program and NSF EEC 1129525 for their support in part of this paper. Additionally we would like to thank Dr. Jeff Morris for his musical expertise, as well as Stephanie Valentine and the Sketch Recognition Lab for their support in the development of this research.

\bibliography{references}
\bibliographystyle{aaai}
\end{document}